\documentclass[pre,onecolumn,amssymb]{revtex4}
\usepackage{epsfig,amsmath}
\begin{document}
\title{Vicious accelerating walkers}
\author{S.-L.-Y. Xu and J. M. Schwarz}

\affiliation{Physics Department, Syracuse University, Syracuse, NY 13244}
\date{\today}
\begin{abstract}
A vicious walker system consists of $N$ random walkers on a line with any two walkers annihilating each other upon meeting. We study a system of $N$ vicious accelerating walkers with the velocity undergoing Gaussian fluctuations, as opposed to the position. We numerically compute the survival probability exponent, $\alpha$, for this system, which characterizes the probability for any two walkers not to meet. For example, for $N=3$, $\alpha=0.71\pm0.01$. Based on our numerical data, we conjecture that $\frac{1}{8}N(N-1)$ is an upper bound on $\alpha$. We also numerically study $N$ vicious Levy flights and find, for instance, for $N=3$ and a Levy index $\mu=1$ that $\alpha=1.31\pm0.03$. Vicious accelerating walkers relate to no-crossing configurations of semiflexible polymer brushes and may prove relevant for a non-Markovian extension of Dyson's Brownian motion model.
\end{abstract}

\maketitle

\section{Introduction} 

Consider $N$ random walkers on a line. The only interaction between the walkers is that if any two walkers meet, they annihilate each other, hence these walkers are deemed vicious.  This system was introduced by M. E. Fisher in his 1983 Boltzmann Medal lecture and was applied to interfacial wetting in $1+1$ dimensions since the interaction of different interfaces (walkers) drives the wetting transition~\cite{Fisher,Fisher.Huse}. In addition, the fermionic nature of vicious walkers provides a Coulomb gas description and, thereby, a link with Gaussian random matrices~\cite{Dyson,Forrester}. More specifically, Baik has proven that a particular limiting conditional distribution of the displacement of the leftmost walker is equivalent to the Tracy-Widom distribution for the Gaussian Orthogonal Ensemble(GOE)~\cite{Baik}. Moreover, vicious walker configurations correspond to directed polymer brushes with the vicious mechanism capturing the non-intersecting property of the polymers~\cite{Essam}. So while the study of vicious walkers has attracted attention from the mathematical physics community, a number of different physical applications also drive its study.

There have been generalizations of vicious walkers to dissimilar walkers~\cite{Gelfand}, walkers with drift~\cite{Wrinkler}, and walkers with external potentials~\cite{Bray}. Here, we introduce a system of $N$ accelerating vicious walkers and study the survival probability, $s(t)$, the probability of having none of the $N$ walkers annihilated up to time $t$. In one-dimension, a randomly accelerating walker, $x(t)$, is defined by

\begin{equation}
\frac{d^2 x(t)}{dt^2}=\eta(t),
\end{equation}
where $\eta(t)$ is Gaussian noise with $\langle \eta(t)\rangle=0$ and $\langle\eta(t)\eta(t')\rangle=2D\delta(t-t')$. The corresponding Fokker-Planck equation is
\begin{equation}
\left[ D\frac{\partial^2}{\partial v^2}-v\frac{\partial}{\partial x}\right]p(x,v,t)=\frac{\partial}{\partial t}p(x,v,t),
\end{equation}
where $p(x,v,t)$ is the probability density distribution for a randomly accelerating walker in one-dimension.

A randomly accelerating walker is presumably the simplest non-Markovian
stochastic process.  Given the fermionic interaction between vicious
accelerating walkers, one can explore the possibility of a non-Markovian
analogue to the Coulomb gas and, hence, a potentially new class of
non-Markovian random matrices. Recently, Fukushima and collaborators have
revisited Dyson's original Brownian motion model for random matrices and found
another non-Markovian stochastic process after generalizing the coefficients
of the matrices~\cite{Fukushima}. In addition, a randomly accelerating walker
appears in the Boltzmann weight of an extensible, semiflexible polymer of
length $L$ with non-zero bending energy~\cite{Burkhardt}. For a given displacement vector $\vec{r}(z)$ from some reference point along a contour length, $z$, the Hamiltonian, $H$, is given by
\begin{equation}
H(\vec{r}(z),\vec{u}(z);z)=\frac{\kappa}{2}\int_0^L{dz}\left(\frac{d^2\vec{r}(z)}{dz^2}\right)^2,
\end{equation}
where $\vec{u}(z)$ denotes the tangent vector and $\kappa$ characterizes the bending rigidity. Mapping the contour length $z$ to time $t$ and the displacement vector $\vec{r}$ to $\vec{x}(t)$, the equation of motion for this system corresponds to a randomly accelerating walker in the corresponding dimension.  Implementing the vicious interaction between $N$ random accelerating walkers, therefore, corresponds to the statistics of non-intersecting semiflexible polymer brushes. In the interest of studying a mixture of flexible and semiflexible polymer brushes, one can also address a mixture of vicious accelerating and random walkers. Please see the appendix for some discussion of $N\le3$ vicious mixed walkers system.

\section{$N\le 3$ vicious accelerating walkers} 

We begin by considering $N=2$ vicious accelerating walkers in one-dimension with equal diffusion constants $D_1=D_2=D$. The two accelerating walkers are governed by the equation,
\begin{eqnarray}
D\left(\frac{\partial^2}{\partial v_1^2}+\frac{\partial^2}{\partial v_2^2}\right )p(x_1,x_2,v_1,v_2,t) \nonumber\\ 
-\left(v_1\frac{\partial}{\partial x_1}+v_2\frac{\partial}{\partial x_2}\right )p(x_1,x_2,v_1,v_2,t)= \nonumber\\
\frac{\partial}{\partial t}p(x_1,x_2,v_1,v_2,t)
\end{eqnarray}
with the initial condition, $
p(x_1,x_2,v_1,v_2,t=0)=\delta(x_1-x_{1,i})\delta(x_2-x_{2,i})\delta(v_1)\delta(v_2)$,
with $x_{1,i}<x_{2,i}$. In addition, to compute the survival probability, $s(t)$, we implement the boundary condition,
\begin{equation}
p(x_1=x_2,v_1>v_2,t)=0
\end{equation}
such that the system ``dies'' when the two walkers meet in space and have ingoing velocities. Note that the boundary condition in $v$ is redundant since, due to the initial positions, the two walkers cannot meet with a relative outgoing velocity.

We then choose the relative coordinate system to reduce the $N=2$ vicious accelerating walkers to one random accelerating walker in the presence of an absorbing wall. This change of variables takes the form, $x=x_1-x_2$ and $v=v_1-v_2$. The Fokker-Planck equation thereby reduces to
\begin{equation}
\left[2D\frac{\partial^2}{\partial v^2}-v\frac{\partial}{\partial x}\right]p(x,v,t)=\frac{\partial}{\partial t}p(x,v,t),
\end{equation}
with the boundary condition, $p(x=0,v>0,t)=0.$

The survival probability for this process is nontrivial in that one cannot invoke the method of images as is done for the ordinary random walker. Using properties of the integral of a Brownian curve, Sinai~\cite{Sinai} proved that the asymptotic survival probability distribution is given by
\begin{equation}
s(t)\sim t^{-1/4}
\end{equation}
at long times. Note that the first-passage time distribution, $f(t)$, where the first passage time is defined by the time at which any of the two walkers meet is given by $f(t)=-\frac{ds(t)}{dt}$ such that $f(t)\sim t^{-5/4}$. In general, if the survival probability distribution is given by $s(t)\sim t^{-\alpha}$ at large times, then $f(t)\sim t^{-\beta}$ with $\beta=\alpha+1$.

A heuristic argument, based on Sinai's approach, for $\alpha=1/4$ was given in Ref.~\cite{Schwarz}. First, a new time counter, $M$, is defined by each ``original'' time the velocity crosses zero. In other words, the original time is now defined by the distribution of first-passage time of a random walk undergoing a Levy flight with Levy index $\mu=1/2$. Moreover, with this new counting, the position variable is also a Levy flight with Levy index $\mu=1/3$.  By invoking the powerful superuniversality of the Sparre-Andersen theorem in one-dimension~\cite{SparreAndersen,Feller,Majumdar}, the first passage time distribution for the position in terms of $M$ is the same as for a random walk, i.e. $M^{-3/2}$. To convert back to the original time, one simply needs to compute the integral
\begin{equation}
f(t)\sim \int \frac{1}{M^{\frac{3}{2}}}\frac{M}{t^{\frac{3}{2}}}\exp(-t^2/M) dM,
\end{equation}
where $\frac{M}{t^{\frac{3}{2}}}\exp(-t^2/M)$ is the limiting distribution for the sum of $M$ $\mu=1/2$ Levy variables. We should also mention that Burkhardt~\cite{Burkhardt} made the use of Marshall-Watson functions\cite{Marshall.Watson} to solve for the Laplace transform version of Eq. 6 with the absorbing boundary condition.

To numerically check for $\alpha=1/4$ result (and other results), we implement
the go-with-the winners algorithm~\cite{Winner}.  We do this because as $N$
increases, the first passage time exponent $\beta$ increases, making it more
difficult to sample the tail of the distribution. The go-with-the winners
algorithm iterates replica systems in parallel. Once at least one pair of
random accelerators have crossed paths in some fraction of the replicas, those
surviving replicas are copied over to the replicas that have already
``died''. We choose that fraction to be one-half such that each copy generated
carries a relative weight of $1/2^c$, where $c$ is the number of copies, which
is then incorporated into the survival probability. The number of replicas range between $1,000$ and $10,000$. The number of runs averaged over range between $10$ and $40$. Fluctuations between the runs are then used for error analysis.

\begin{figure}[t]

\epsfig{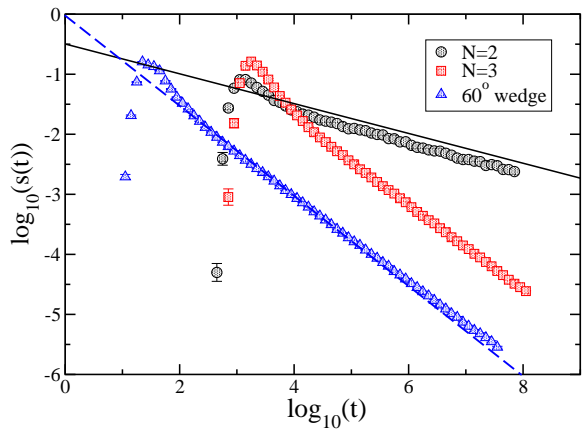}

\caption{Log-log plot of survival probability distribution versus time for $N=2$ and $N=3$ vicious accelerating walkers in line and for a single accelerating walker in a 60$^\circ$ wedge geometry. The line denotes a survival probability exponent of $1/4$, while the dashed line denotes a survival probability exponent of $3/4$. }

\end{figure}

Having calibrated our results for the $N=2$ case, we now consider $N=3$ vicious accelerating walkers with equal diffusion constants. The Fokker-Planck equation is given by
\begin{equation}
\sum_{j=1}^3\left[D\left(\frac{\partial^2}{\partial v_j^2}\right ) 
-\left(v_j\frac{\partial}{\partial x_j}\right)\right]p(X,V,t)
=\frac{\partial}{\partial t}p(X,V,t),
\end{equation}
where $X=\{x_1,x_2,x_3\}$, and $V=\{v_1,v_2,v_3\}$. 

We apply change of variables $v=v_1-v_2$, $u=v_2-v_3$, $x=x_1-x_2$, $y=x_2-x_3$ such that the LHS operator for two relative coordinates becomes
\begin{equation}
2D\left(\frac{\partial^2}{\partial v^2}+\frac{\partial^2}{\partial u^2}-\frac{\partial^2}{\partial v\partial u}\right)-\left(v\frac{\partial}{\partial x}+u\frac{\partial}{\partial y}\right)
\end{equation}
with the absorbing boundary at $x=0$, $y=0$, $u>0$ and $v>0$. Again, here, boundary conditions on the velocities are redundant. To remove the coupled term in $u$ and $v$, we perform another set of linear transforms, $u=\frac{l-q/\sqrt{3}}{2}, v=\frac{-l-q/\sqrt{3}}{2}, x=\frac{-z-w/\sqrt{3}}{2},$ and $y=\frac{z-w/\sqrt{3}}{2}$ to obtain
\begin{equation}
6D\left(\frac{\partial^2}{\partial q^2}+\frac{\partial^2}{\partial l^2}\right)-\left(q\frac{\partial}{\partial w}+l\frac{\partial}{\partial z}\right)
\end{equation}
with absorbing boundaries, $z=\pm w/\sqrt{3}$, i.e. a $60^{\circ}$ wedge in the $z-w$ plane.

We have reduced three vicious accelerating walkers in one-dimension to one accelerating walker in two-dimensions in a wedge geometry.  For comparative purposes, let us review the ordinary random walker in a wedge~\cite{Redner.Book}. Using the conformal mapping $z'=z^{\pi/\theta}$, one can map the wedge geometry to the upper-half plane. Then, the motion in $x$-direction becomes unbounded and in the $y$-direction it is the simple situation of a one-dimensional walker with an absorbing boundary condition. The survival probability distribution asymptotically in the upper-half plane is $s(t)\sim t^{-1/2}$. After an inverse conformal mapping, one arrives at the survival probability distribution of $s(t)\sim t^{-\pi/2\theta}$.






While there is currently no analytical solution for the survival probability distribution for an accelerating walker in a $60^{\circ}$ wedge geometry, for a $90^{\circ}$ wedge, the Sparre-Andersen theorem can be invoked for the two indepedent directions to arrive at an asymptotic power-law survival probability distribution with $\alpha=\frac{1}{2}$. In addition, for a $180^{\circ}$ wedge, $\alpha=\frac{1}{4}$. While the Sparre-Andersen theorem is quite powerful and can easily be extended to as many independent dimensions as needed, the $60^{\circ}$ wedge geometry couples the two directions (for $N=3$, at least) and, therefore, the Sparre-Andersen theorem, as it stands, cannot be invoked.

However, to smoothly interpolate between the $90^{\circ}$ and $180^{\circ}$
cases, we conjecture that for other wedge angles, the survival probability
distribution also asymptotes to a power-law with survival probability
exponent, $\alpha$, with $\alpha$ decreasing continuously as the wedge angle
increases.  To this conjecture, we resort to numerical simulation of both the
wedge geometry for one accelerating walker and the line geometry for three
accelerating walkers. The result is presented in Fig. 1. We measure a survival
probability exponent of $\alpha=0.71\pm0.01$ for $N=3$ vicious accelerating
walkers in a line, which agrees with the $60^{\circ}$ wedge geometry result
(with essentially the same error bar).  Also, referring back to the accelerating walker-Levy flight mapping implemented to demonstrate the $\alpha=\frac{1}{4}$ exponent for an absorbing accelerating walker in one-dimension, simulating a $\mu=1/3$ Levy flight in a $60^{\circ}$ wedge geometry with the opening angle, $\theta$, between $0^{\circ}$ and $60^{\circ}$, yields $\alpha=0.71\pm0.03$.

Figure 2 tests our conjecture that $\alpha$ continuously decreases with increasing wedge angle. Both the numerical values of $\alpha$ for the $90^{\circ}$ and $180^{\circ}$ wedges agree well with their analytical counterparts.  For comparison purposes, we have also plotted the curve $\alpha=\pi/4\theta$, which agrees with the two analytical solutions and can be viewed as a trivial extension of the random walker solution.  While the agreement looks reasonable at larger angles, the deviation becomes more apparent at smaller angles. It also appears that the divergence in $\alpha$ as $\theta$ decreases to zero is slower than $1/\theta$. 

\begin{figure}[t]

\epsfig{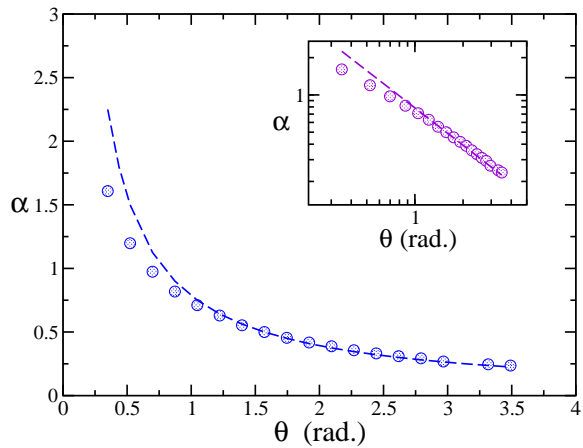}

\caption{Survival probability exponent for a two dimensional accelerating walker in a wedge geometry as a function of the opening angle.  The line denotes $\alpha=\frac{180}{4\theta}$. Inset: Log-log plot to emphasize the difference between the data and $\alpha=\frac{180}{4\theta}$. The error bars are smaller than the symbols.}

\end{figure}

\section{$N>3$ vicious accelerating walkers} 

Now, we numerically address $N>3$
vicious accelerating walkers. To compare with ordinary vicious walkers, based
on the method of images, Fisher~\cite{Fisher,Fisher.Huse} considered one
compound walker in $N$ dimensions that cannot cross any of the
$x_1=x_2,x_2=x_3,...,$ or $x_{N-1}=x_N$ linear manifolds. Using the method of
images, as long as the initial positive and negative weights (corresponding to
unrestricted positive and negative walkers) are chosen such that probability
distribution is antisymmetric under reflection within each linear manifold,
then the distribution will satisfy the absorbing boundary conditions for all
future times. These weights can be represented as a Vandemond determinant,
which can be factorized to yield a product of $\frac{1}{2}N(N-1)$ pairings
such that $\alpha(N)=N(N-1)/4$, where $\alpha(2)=1/2$. Previous and the current simulations verify this prediction. See Figure 3. By the same argument, if the survival exponent for two (one pair of) vicious accelerating walkers is $1/4$, the survival exponent for $N$ vicious accelerating walkers system should be $N(N-1)/8$. However, for vicious accelerating walkers, the method of images fails.

Based on the $N=2$ and $N=3$ results, combined with the fact that $N$ vicious accelerating walkers can be mapped to one accelerating walker in $N-1$ dimensions in an unbounded domain, we expect the power-law survival probability distribution extends to $N>3$. In Figure 3 we present the simulation results of survival probability exponents for vicious accelerating walkers system up to $N=10$, as well as the too trivial prediction, $\alpha=N(N-1)/8$. The deviation is apparent at large $N$. We find that $N(N-1)/8$ is clearly an upper bound for the measured $N$.  The non-Markovian nature of the accelerating walker enables the system to survive longer and in a way that cannot be accounted for by individual pairings.

\begin{figure}[t]

\epsfig{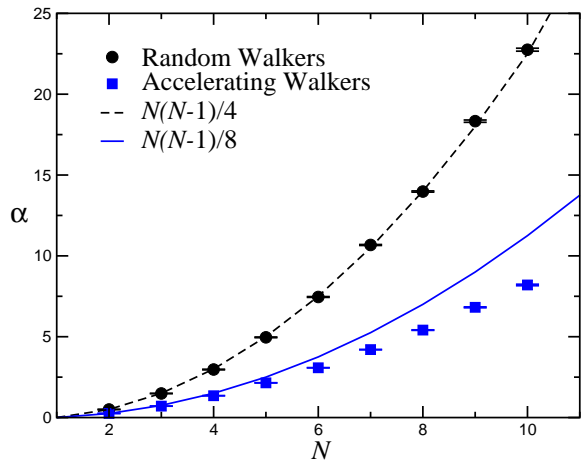}

\caption{Survival probability exponent $\alpha$ for vicious accelerating walkers (solid squares) and vicious Gaussian walkers (solid circles) systems up to $N=10$. The simulation results for vicious Gaussian walkers agree very well with theory prediction $\alpha=N(N-1)/4$ (black curve). However, the results for vicious accelerating walkers deviate from a method of images prediction of $\alpha=N(N-1)/8$ (blue dashed curve). }

\end{figure}

\section{Vicious Levy flights} 

As mentioned previously, vicious Gaussian walkers
problem is closely related to the Gaussian random matrix theory. A matrix with
random entries falls into this category as long as the entries are independent
and identically distributed (iid) variables with a finite second moment of the
corresponding distribution. A generalization of the random Gaussian matrix is
the random Levy matrix~\cite{RLM}, where the entries are drawn from a broader
distribution, namely a Levy distribution. The most important characteristic of
Levy distribution is a heavy power-law tail step-size $S$ distribution,
$P(S)\sim S^{-1-\mu}$ for large $S$. When $\mu\ge 2$, the variance of the
distribution is finite and, hence, the central limit theorem holds for the
distribution of the sum of independently drawn Levy variables. Similarly, the
random Levy matrices reduce to random Gaussian matrices. However in the regime
$\mu<2$, the variance of the distribution diverges and hence random Levy
matrices behave qualitatively different from random Gaussian matrices. For
example, the famous Wigner-Dyson semicircular law is replaced with a density
of states that extends over the entire eigenvalue axis~\cite{RLM}.  

Inspired by the connection between vicious Gaussian walkers and random
Gaussian matrices as well as the connection between Levy flights and random
accelerating walkers, we study the problem of vicious Levy flights. 
Given $N$ Levy flights in one-dimension, we define the vicious interaction
between pairs. Because Levy flights are nonlocal, two Levy flights jump over
each other without meeting at some exact point.  Hence, there could be two
ways to define the vicious interaction, to prohibit the jump-overs or to allow
for jump-overs and the annihilation occurs upon intersection within some
range, irrespective of the ordering. The latter has been recently studied\cite{VLF}. However, we are more interested in the former case, in which the set of Levy flights annihilate whenever a crossing occurs. In other words, a surviving system strictly maintains the initial ordering of all flights, which is the same as for the vicious Gaussian walkers.

\begin{figure}[t]
\vspace{0.5cm}
\epsfig{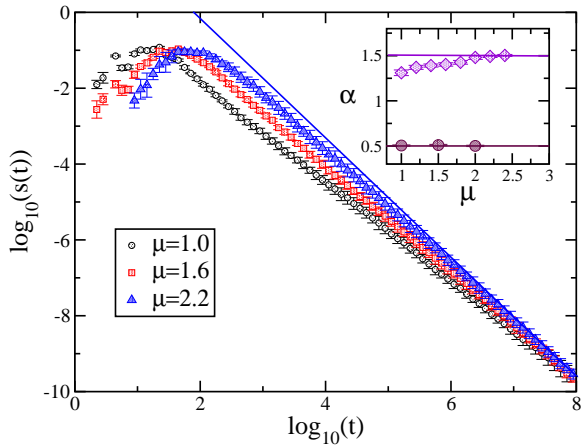}

\caption{Log-log plot of the the survival probability distribution for $N=3$ vicious Levy flights for different values of the Levy index, $\mu$. The curve denotes a survival probability exponent of $\alpha=\frac{3}{2}$. Inset: The survival probability exponent for $N=2$ and $N=3$ vicious Levy flights as a function of the Levy index.  For $\mu>2$, we obtain the vicious random walker results. }

\end{figure}

The Fokker-Planck equation for a system of $N$ one-dimensional vicious Levy flights is described as
\begin{equation}
\sum_{j=1}^{N}\frac{\partial^\mu}{\partial |x_j|^\mu}p(X,t)=\frac{\partial}{\partial t}p(X,t),
\end{equation}
where the normal Laplacian is replaced by the Riesz-Feller derivative of fractional order $2>\mu>0$~\cite{Podlubny,Samko}. This derivative has an integral representation, which more easily reveals its nonlocal nature. The initial condition is still $p(X,t=0)=\prod_{j=1}^N\delta(x-x_{j,i})$, with $x_{j,i}<x_{k,i}$ for all $j<k$. The boundary condition for the non-crossing vicious interaction as we described above is then $p({x_j},t)=0$, if $x_j\ge x_k$ for any $j<k$.

The $N=2$ case is, again, equivalent to the first-passage problem of a single
Levy flight via a transformation to relative coordinates (and
integrating out the center of mass coordinate). The only
difference with a random walker is that the absorbing boundary condition at
the origin has to be modified to an absorbing region occuping the positive
x-axis to preserve the non-crossing property. The first-passage property of a
Levy flight is governed by the Sparre-Andersen
theorem\cite{SparreAndersen,Feller,Majumdar}, which implies that the first-passage time
distribution for any symmetric step size distribution in one-dimension
asymptotes to the same as that of a Gaussian walker.  Thus, the survival probability exponent for $N=2$ is $\alpha=1/2$ independent of $\mu$. We verify this result in our simulation. See Figure 4.  Note that this result is very different from the result obtained in Ref.~\cite{VLF} where $\alpha$ depends on $\mu$ for $N=2$ and higher.

We also simulate $N=3$ vicious Levy flights.  Because of the linearity of
fractional derivatives, the mapping of two vicious Levy flights to a single
Levy flights in an absorbing plane holds. However, due to the lack of
rotational invariance of the Riesz-Feller derivative, the wedge mapping that
holds for vicious walkers and now for vicious accelerating walkers, does not
apply to vicious Levy flights. In order to make progress, since for $N=2$
there exists a power-law distribution, we conjecture that the survival
probability distribution scales as a power-law at long times for $N>2$ and
measure $\alpha$. Figure 4 plots the survival probability exponents for $N=3$
vicious Levy flights for several different Levy indices. For $N=2$ all values
of $\mu$ yield the same survival probability exponent of 1/2, in agreement
with the Sparre-Andersen theorem. However, the $N=3$ exponents appear to vary
with $\mu$.  For instance, for $\mu=1$, $\alpha=1.31\pm0.03$. While the 0.19
difference between $\mu=1.0$ and $\mu=2$ is small, the difference grows with
$N$.  For example, for $N=4$ and $\mu=1$, $\alpha=2.3\pm0.1$ and for $\mu=2$,
$\alpha=2.91\pm0.09$. Based on this data, we speculate that for $N>2$,
$\alpha$ depends on $\mu$.

A few comments on the technical aspects of the simulations are in order. We implement an upper cut-off on the Levy steps so that at long enough times, the survival probability distribution approaches the random walker result~\cite{Mantegna}.  The convergence also depends on the Levy index. For example, for $\mu=1$ and stepsize cut-off $S_c=10-100$, the convergence to the random walker result is fast such that the asymptote to a power-law beyond $t\approx 10^2$ is in agreement with the random walker result to within one standard deviation. In contrast for $\mu=1.6$, $S_c= 10^9$, and time scales beyond $t\approx 10^8$, convergence to the random walker result is observed. Secondly, for $\mu=1$ we also generated Cauchy distributed numbers directly and found good agreement with the power-law generated $\mu=1$ result.     

\section{Conclusion} 

To summarize, we have generalized the vicious walker problem
 in two different ways: (1) vicious accelerating walkers and (2) vicious Levy
 flights as defined by non-crossing. For both generalizations, the typical
 analytical technique of the method of images fails. Analytical results for
 $N=2$ are readily obtainable since both problems can be mapped to the first
 passage problem of a single accelerating walker or Levy flight with the
 appropriate absorbing boundary or region. We demonstrate that the $N$ vicious
 walker mapping to one walker in $N-1$ dimensions in a wedge geometry
 generalizes to vicious accelerating walkers. We also conjecture, based on our
 numerical data, that there exists an upper bound on the survival probability
 exponent of $\alpha=\frac{1}{8}N(N-1)$ for $N$ vicious accelerating
 walkers. An analytical calculation for the $N=3$ case corresponding to one
 acclerating walker in a two-dimensional wedge geometry would be the next
 logical step.  The heuristic argument for the absorbing accelerating walker
 in one-dimension using a new time counter and Levy flights may eventually
 become useful to analyze the two-dimensional wedge problem. There also exists a recent numerical result for the
 survival probability distribution for the two-dimensional Fractional Browian
 motion process, originally introduced by Kolmogorov~\cite{Kolmogorov}, in a
 wedge~\cite{Jeon}. We anticipate more study of non-Markovian processes in
 dimensions higher than unity in the near future. Indeed, a non-Markovian
 extension of Dyson's Brownian motion model to, for example, include inertia,
 may be related to $N$ vicious accelerating walkers to arrive at a new class
 of random matrices.  It may also be interesting to investigate other ordering
 problems of randomly accelerating walkers on a line such as the Gaussian
 equivalent of the ''leader'' and the ''laggard'' problem~\cite{benAvraham}.

Finally, given our numerical results, we speculate that the survival probability exponent for $N$ vicious Levy flights (as defined by no-crossing) depends on $\mu$ for fixed $N>2$. We also refer to a new result where vicious Levy flights are defined as annihilating when any two Levy flights come within some range of each other and $\alpha$ depends on $\mu$ even for $N=2$~\cite{VLF}.  While the survival probability exponent in one-dimension is independent of the Levy index, as a consequence of the powerful Sparre-Andersen theorem, we anticipate that this superuniversality may be broken in dimensions higher than unity and the universality of each Levy index becomes exposed. In light of our results, a higher dimensional generalization (or modification) of the Sparre-Andersen theorem should be on the forefront of at least several statistical physicists and mathematicians minds. 

JMS would like to acknowledge M. Jeng for performing some preliminary simulations on no-crossing vicious Levy flights, Eli Hawkins for helpful discussion, and The Aspen Center of Physics where some of this work was conducted. JMS is supported by NSF-DMR-0645373.


\appendix*
\section{$N\le3$ mixed case}

Consider a mixture of vicious accelerating and random walkers---an example of a vicious walkers system with dissimilar members.  For one random walker and one random accelerator, the random walker can be approximated as an absorbing boundary at rest since the displacement of a random walker scales as $x(t)\sim t^{1/2}$, while for a random accelerating walker, the velocity scales as $v(t)\sim t^{1/2}$ such that the displacement scales as $x(t)\sim t^{3/2}$. Thus, in long time limit, the displacement of a normal random walker is negligible compared with that of a random accelerator. In other words, one can simply take the normal random walker as a fixed absorbing wall to fulfill the vicious mechanism. Simulations of one random walker and one accelerating walker that annihilate upon crossing verify this.

Now to address the $N=3$ mixed case. We use the ''A'' to denote an
accelerating walker and ''R'' to denote a normal one in the present
discussion. The two pure states ''RRR'' and ''AAA'' are trivial with the former following Fisher's general solution and latter demonstrated by our numerical simulation described earlier in the paper. The combination with two normal walkers and one accelerator can have two forms, namely ''RRA'' and ''RAR''. The former can be treated as one absorbing random walker and one absorbing accelerating acting independently such that $\alpha=3/4$. The latter is an accelerating walker sandwiched between two random walkers, i.e. a bounded domain, and hence the survival probability decays instead exponentially in time. A detailed analysis can be found in Ref.\cite{Bicout}. Simulations support both results.  See Figure 5. For "ARA", there is an absorbing wall in between two accelerating walkers, i.e. a pair of decoupled accelerator-wall systems, hence $\alpha=1/2$. The last possible configuration, ''AAR'', is nontrivial in that it can be reduced to one accelerating walker in a $45^{\circ}$ wedge geometry, just as two vicious random walkers in the presence of an absorbing wall can be mapped to one random walker in a $45^{\circ}$ wedge geometry.  We measure a nontrivial survival probability exponent of $\alpha=0.89\pm0.01$ for this case, which is consistent with our wedge measurements. 

\begin{figure}[t]
\epsfig{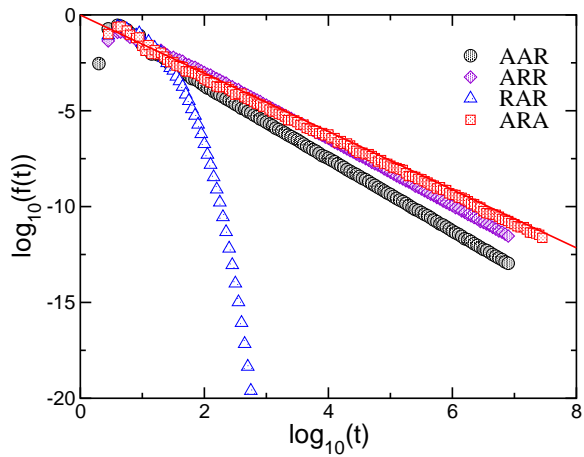}
\caption{Log-log plot of the survival probability distribution for the $N=3$ mixed cases of vicious accelerating walkers (A) and vicious random walkers (R). The line denotes a first passage time probability exponent of $\beta=\frac{3}{2}$.}
\end{figure}

\end{document}